\documentstyle[preprint,aps,epsf]{revtex}

\draft
\title{\bf Lorentz Integral Transform for Inclusive and Exclusive Cross 
Sections with the Lanczos Method}
\author{Mario Andrea Marchisio$^1$, Nir Barnea$^2$, Winfried Leidemann$^1$ 
and Giuseppina Orlandini$^1$}

\address{$^1$ Dipartimento di Fisica, Universit\`a di Trento 
and INFN (Gruppo collegato di Trento) \\ I-38050 Povo (Trento), 
Italy\\ 
$^2$The Racah Institute of Physics, The Hebrew University,
91904 Jerusalem, Israel}
\date{\today}

\begin{document}
\maketitle
\begin{abstract}
\noindent
The Lorentz Integral Transform (LIT) method is reformulated via the Lanczos 
algorithm both for inclusive and exclusive reactions. The new 
technique is tested for the total photoabsorption cross 
section of $^{3}$H and $^{4}$He. Due to the rapid convergence of the algorithm 
one has a decrease in cpu time by two orders of magnitude, 
but at the same time an excellent agreement with the results of a conventional
LIT calculation. The present work opens up the possibility of ab 
initio calculations for inclusive and exclusive processes for A$\ge$6 with 
inclusion of complete final state interactions.
\end{abstract}

\bigskip
\noindent {\footnotesize PACS: 21.45.+v; 25.20.Dc; 25.30.Fj }

\vfill\eject
\section{Introduction}
The study of reaction cross sections is an important tool to reveal the 
dynamics of particle systems. For systems with a small number of
particles one aims at microscopic calculations trying 
to take into account all relevant degrees of freedom of the considered 
process. 
However, calculations via the classical approach, where one employs initial 
and continuum state wave functions, are very difficult to perform for 
reactions at 
energies beyond a possible three-body breakup threshold. On the other hand the 
calculation can be carried out in an alternative way, namely via the Lorentz 
Integral Transform (LIT) method \cite{ELO94}. This method allows ab initio 
calculations which take into account final state interactions  
correctly, without explicit knowledge of the complicated continuum wave 
functions. Various applications of this method for reactions of nuclei up to 
four nucleons can be found in the literature (see e.g. 
\cite{ELO94,Sara,ELO97a,ELO97b,ELO97c,ELO99,ELOT00}).

The LIT is expressed in terms of Lorentz states $\widetilde \Psi$, which have 
bound state like asymptotic boundary conditions that are much simpler to treat 
than the boundary conditions for a continuum state and thus $\widetilde\Psi$ 
can be calculated with similar methods as a bound state wave function.
Common to many bound state methods is an expansion of the wave function
on some basis set. In fact in the above mentioned LIT calculations 
an expansion on the hyperspherical harmonics basis has been employed for the 
reactions of nuclei with  $ A \geq 3$ (with the exception of Ref. 
\cite{Sara}). 
As the Hilbert space increases with the number of particles the 
calculations face larger and larger hamiltonian matrices, which have to be 
inverted in order to solve the equations associated with the LIT method. 
Such an inversion is time consuming and limits the number of 
possible basis states used in the calculation. The aim of the 
present paper is to point out that the inversion can be avoided. Inspired by 
recent condensed 
matter calculations \cite{dagotto,hallberg,kuehner} we have adapted the 
Lanczos method for the calculation of the LIT. In this way we open the 
possibility to extend the LIT method to problems, where a rather large number 
of basis functions is needed, e.g., calculations of reactions of the $\alpha$ 
particle with realistic interactions or calculations of reactions  of systems 
with more than four particles. 

The outline of the paper is as follows.
In Sections II/III we review the LIT method for inclusive/exclusive 
reactions briefly and discuss the corresponding implementation of the Lanczos 
technique. Various checks of the calculation are discussed in Section IV, 
which also contains a short summary at the end.

\section{Inclusive Processes}

As pointed out for the case of the Stieltjes transform \cite{Efros85} 
the integral transform method can be used for the calculation of a variety of 
reaction cross sections. In this and in the following section we restrict the 
discussion to inclusive and exclusive cross sections due to an external probe, 
though the approach can also be applied to other 
reaction cross sections, e.g., to the elastic scattering of particle systems. 

In order to determine inclusive cross sections due to external probes
one has to calculate various dynamical functions 
\begin{equation} \label{effe}
  F_{kk'}(\omega,q)=\int d\psi_f \langle \psi_0 |{\hat{O}}^\dagger_k(q)
   |\psi_f\rangle\langle \psi_f |{\hat{O}}_{k'}(q)
   |\psi_0\rangle \delta (E_f -E_0 -\omega )\,,
\end{equation}
where $\omega$ and $q$ are energy and momentum transfer;
$\psi_0$ and $\psi_f$ denote ground and final state wave functions
of the considered system with energies $E_0$ and $E_f$, respectively, 
while the various ${\hat{O}}_k$ are excitation operators inducing different 
types of transitions. 

For few-body reactions with $A > 2$ one faces very often the problem that
$F_{kk'}(\omega,q) $ cannot be calculated exactly, since the microscopic 
calculation of $\psi_f$ is too complicated. However, via the LIT approach the 
problem can be reformulated in such a way that the knowledge of 
$\psi_f$ is not 
necessary \cite{ELO94}. To this end the LIT of the response function is 
introduced
\begin{equation} \label{lorenzo}
  L_{kk'}(\sigma )=\int d\omega \frac{F_{kk'}(\omega)}{(\omega -\sigma_R)
               ^2+\sigma_I^2}\,,
\end{equation}   
where 
\begin{equation} \label{sigma}
  \sigma=\sigma_R+i\sigma_I \hspace{10mm} \sigma_R,\sigma_I>0  \mbox{.}
\end{equation}
In (\ref{lorenzo}), as well as in the following equations, we 
suppress the dependence on $q$ of $L_{kk'}(\sigma,q)$ and $F_{kk'}(\omega,q)$.
     
The LIT method proceeds in two steps. First $L_{kk'}(\sigma)$ is evaluated
in a direct way, which does not require the knowledge of $F_{kk'}$, 
and then in a 
second step the dynamical function is obtained from an inversion of the LIT. 
The direct calculation of
$L_{kk'}(\sigma)$ is  outlined in the following. Starting from the definition
 of the 
LIT in (\ref{lorenzo}), substituting for $F_{kk'}(\omega)$ the expression in 
(\ref{effe}) and  using the completeness relation of the 
Hamiltonian eigenstates,
\begin{equation} \label{compl1}
   \int df |\psi_f \rangle \langle \psi_f | = 1 \mbox{,}
\end{equation}
one finds
\begin{equation}\label{lorenzog}
  L_{kk'}(\sigma)=\langle \psi_0 | {\hat{O}}^{\dagger}_k\frac{1}{\hat
              {H}-E_0-\sigma_R-i\sigma_I}\frac{1}{\hat{H}-E_0-\sigma_R+i
              \sigma_I}\hat{O}_{k'}|\psi_0 \rangle \mbox{.}
\end{equation}
It is evident that the solutions of the equations 
\begin{equation} \label{psi1}
  (\hat{H}-E_0-\sigma_R+i\sigma_I )|\widetilde{\psi}_l\rangle =\hat
   {O}_l | \psi_0\rangle \,;\,\,\,\,\,\,\,\,\,\, l=k,k'
\end{equation}
lead directly to the searched transform:
\begin{equation} \label{elle}
  L_{kk'}(\sigma )=\langle \widetilde{\psi}_k |\widetilde{\psi}_{k'} 
  \rangle \mbox{.} 
\end{equation}
Physical solutions for the Lorentz states $|\widetilde{\psi}_l \rangle$ of 
(\ref{psi1}) have asymptotic boundary conditions like a bound state. 
Moreover the solutions are unique, since due to the hermiticity 
of $\hat{H}$ the homogeneous equation 
\begin{equation}
\label{psi1a}
(\hat{H}-E_0-\sigma_R+i\sigma_I ) |\widetilde{\psi}_l\rangle  =0
\end{equation}
has  only the trivial solution $|\widetilde{\psi}_l\rangle =0$. 

The inversion of the calculated $L_{kk'}(\sigma)$ leads to the dynamical
functions $F_{kk'}(\omega)$. 
For details on the inversion procedure see, e.g., Ref. \cite{ELO99}. 
The LIT method leads to reliable results as shown in test calculations for
two- and three-body systems \cite{ELO94,Sara}.

If one solves (\ref{psi1}) expanding $|\widetilde{\psi}_l \rangle$ on a set
of basis functions $|\mu \rangle$ one has to invert the corresponding 
hamiltonian
matrix $H_{\mu\, \mu'}$. For a large number of basis functions 
this is a time consuming calculation, moreover, for dense matrices 
this task rapidly becomes prohibitive.
As we will show, the Lanczos method can help to overcome this problem.

Basically the Lanczos method \cite{lanczos} is used to 
{\it{tridiagonalize}} matrices.
If one is interested in calculating the ground state energy 
of an hamiltonian matrix $H_{\mu\, \mu'}$ one has first to choose a starter,
a vector $|\phi_0 \rangle$, which must have a non-zero
overlap with the ground state $|\psi_0 \rangle$. One can then 
build the Lanczos orthonormal basis $\{|\phi_i 
\rangle \mbox{,}i=0,\ldots \mbox{,} n\}$ by applying the 
Lanczos algorithm recursively    
\begin{equation} \label{lana}
   b_{i+1}|\phi_{i+1}\rangle =\hat{H}|\phi_i\rangle - 
   a_i|\phi_i\rangle -b_i|\phi_{i-1}\rangle \,,
\end{equation}
where
\begin{equation}
   |\phi_{-1}\rangle =0 \mbox{,}\,\,\,\,\,\,
   \langle \phi_i |\phi_j \rangle =\delta_{ij} \mbox{.}
\end{equation}
The Lanczos coefficients $a_i$ and $b_i$ are defined as
\begin{equation}
   a_i=\langle \phi_i |\hat{H}|\phi_i \rangle \mbox{,}\,\,\,\,\,\, 
   b_i=\parallel b_i|\phi_i \rangle \parallel \mbox{,}\,\,\,\,\,\,
   b_0=0 \mbox{.} 
\end{equation}
If one applies $M$ Lanczos steps to an $M\times M$ matrix,
then the original 
matrix is reduced into a tridiagonal form. However, the power of the Lanczos
algorithm lays in the fact that even after $ m << M$ steps the eigenvalues 
of the $m \times m $ tridiagonal Lanczos matrix are good approximations of the
extreme eigenvalues of $H$. These eigenvalues converge rapidly with increasing 
number of Lanczos steps. 
Working with finite precision, it is necessary to 
recall that at each step there are 
round-off errors so that after a certain number of steps
the Lanczos vectors loose their orthogonality. In order to 
obtain accurate results this loss has to remain small, otherwise
one has to apply an additional reorthogonalization procedure.

In the following we show how the Lanczos algorithm can be applied to the 
LIT method for inclusive processes. To this end we rewrite the LIT in the
following form
\begin{equation} \label{lorelan1}
\frac{\sigma_I}{\pi}\,L_{kk'}(\sigma)=-\frac{1}{\pi} \mbox{Im} 
\left\{ \langle \psi_0 |\hat{O}_k^{\dagger}\frac{1}{\sigma_R+i\sigma_I+E_0-
\hat{H}} \hat{O}_{k'} |\psi_0 \rangle \right\}\mbox{.}
\end{equation}
A similar relation connects the dynamical functions $F_{kk'}$ to 
the Green's function
\begin{equation} \label{lorelan2}
F_{kk'}(\omega)=-\frac{1}{\pi} \mbox{Im} \left\{\lim_{\eta\to 0} 
G(\omega + i \eta + E_0)\right\}\,;\,\,\,\,\,\,\,\,\,G(z)=\langle \psi_0 
  |\hat{O}_k^{\dagger}\frac{1}{z-\hat{H}} 
  \hat{O}_{k'} |\psi_0 \rangle\,,
\end{equation}  
provided that $z = \omega + i \eta$ is replaced by $\sigma_R + i \sigma_I$.
This is not surprising since for $\sigma_I \rightarrow 0$ the properly
normalized Lorentzian kernel
tends to  $\delta(\omega-\sigma_R)$ and thus $L_{kk'}(\sigma_R) \sigma_I/\pi 
\rightarrow F_{kk'}(\sigma_R)$. In condensed 
matter calculations \cite{dagotto,hallberg,kuehner} the Lanczos algorithm has
been applied to the calculation of the Green function with a small value of 
$\eta$, and its imaginary part has been interpreted as $F_{kk'}(\omega)$
directly. This can be done if the spectrum is discrete (or discretized) and
$\eta$ is sufficiently small. In our case we have a genuine continuum
problem and we want to avoid any discretization, therefore we calculate
$L_{kk'}(\sigma_R)$ in the same way i.e. with finite $\sigma_I$ using the 
Lanczos algorithm, but then we antitransform $L_{kk'}(\sigma_R)$ in order to 
obtain $F_{kk'}(\omega)$.

Choosing 
\begin{equation} \label{start}
  |\phi_0\rangle =\frac{\hat{O}|\psi_0\rangle }{\sqrt{\langle 
  \psi_0|\hat{O}^{\dagger}\hat{O}|\psi_0 \rangle }}
\end{equation}
as starting vector for the Lanczos basis and setting 
$z=E_0+\sigma_R+i\sigma_I$ one finds 
\begin{equation} \label{loreins}
  L(\sigma)=\frac{1}{\sigma_I}  
            \langle \psi_0 | \hat{O}^{\dagger}\hat{O}|\psi_0\rangle 
            \mbox{Im} \{\langle \phi_0 |
                      \frac{1}{z-\hat{H}}|\phi_0\rangle\} \mbox{}
\end{equation}
showing that the LIT depends on the matrix element
\begin{equation}
  x_{00}= \langle \phi_0 |\frac{1}{z-\hat{H}}|\phi_0\rangle \,.
\end{equation}
One can calculate $x_{00}$ applying Cramer's rule to the solution of the 
linear system 
\begin{equation}
  \sum_n (z-\hat{H})_{mn}x_{n0}=\delta_{m0}
\end{equation}
that arises from the expansion of the identity 
\begin{equation}
  (z-\hat{H})(z-\hat{H})^{-1}=I 
\end{equation}
over the Lanczos basis \{$|\phi_i\rangle \mbox{,} i=0\mbox{,}\ldots\mbox{,} n\}
$ \cite{dagotto,fulde}.
In this way one is able to write $x_{00}$ as a continued fraction containing 
the Lanczos coefficients $a_i$ and $b_i$, 
\begin{equation} \label{conti}
  x_{00}=\frac{1}{(z-a_{0})-\frac{b^{2}_{1}}{(z-a_{1})-
                  \frac{b^{2}_{2}}{(z-a_{2})-b^{2}_{3}....}}} \,,
\end{equation}
and thus also the LIT becomes a function of the Lanczos coefficients
\begin{equation} \label{lorelan3}
L(\sigma )=\frac{1}{\sigma _{I}}\mbox{Im}\left \{ -\frac
          {\langle \psi _{0}|\hat{O}^{\dagger}\hat{O}| \psi _{0}\rangle
           }{(z-a_{0})-\frac{b^{2}_{1}}{(z-a_{1})-\frac{b^{2}_{2}}{
          (z-a_{2})-b^{2}_{3}....}}}\right \} \,.
\end{equation}
This illustrates that the Lanczos method allows to determine $L(\sigma )$ 
without solving the differential equation (\ref{psi1}), i.e. without inverting 
the hamiltonian matrix.

\section{Exclusive Processes}
Different from an inclusive reaction where the final state of the studied 
particle system is unobserved, for an exclusive process one has to determine 
the cross section of a specific breakup channel for specific angles of the 
outgoing fragments.
For instance, the unpolarized cross section for a two-body
breakup induced by an 
external probe is given by
\begin{equation}
\label{cross}
\frac{d\sigma}{d\Omega_f} = \sum_{kk'} c_{kk'}(\omega,q,\phi_f) 
f_{kk'}(\omega,q,\theta_f) \,,
\end{equation}
where $\Omega_f=(\theta_f,\phi_f)$ is the spherical angle of one of the 
outgoing fragments. In (\ref{cross}) the
$c_{kk'}$ are known kinematical functions, while the dynamics of the particle
system is contained in the structure functions 
\begin{equation}
f_{kk'}(\omega,q,\theta_f) = \sum_{\alpha m} T^*_{\alpha k m}(E_f)  
T_{\alpha k' m}(E_f) \,.
\end{equation}
The above transition matrix elements are defined as follows
\begin{equation}\label{ti1}
  T_{\alpha k m}(E_f) = 
  \langle \psi_{f,\alpha} | {\hat O}_{k}(q) | \psi_{0,m} \rangle \,,
\end{equation} 
where $\alpha$ stands for a set of quantum numbers of the final state function
$\psi_f$ with energy $E_f$ in the specific breakup channel and $m$ is the 
projection of the total angular momentum of the initial state wave 
function $\psi_0$.

Different from the inclusive cross section, where an integral method  leads to 
the direct determination of the dynamical function $F_{kk'}$,
one has to determine any single 
transition matrix element $T_{\alpha k m}(E_f)$  in order to calculate 
the exclusive cross section \cite{Efros85}.

We illustrate the derivation of the integral 
transform method for the exclusive case restricting ourselves to a breakup of 
the A-body system in two fragments. In this case the final state of the system 
can be written using the Lippmann-Schwinger equation
\begin{equation}
\label{lippmann}
\langle \psi_{f,\alpha}| =\langle \phi_{\alpha}^{PW}| +
\langle \phi_{\alpha}^{PW}|\hat{V}_{12}\frac{1}{E_f+i\epsilon -\hat{H}} \,\,,
\end{equation}
where $|\phi_{\alpha}^{PW}\rangle$ describes the plane wave relative motion of 
the two fragments with the same set of quantum numbers $\alpha$ as 
$\psi_{f,\alpha}$, while $\hat{V}_{12}$ denotes the interaction 
of the constituents
of one fragment with the constituents of the other fragment. In coordinate
space the plane wave has the following form
\begin{equation} \label{plane}
  \phi_{\alpha}^{PW}(\vec r\,)=
         {\mathcal{A}}\psi_1 \psi_{2} \frac{e^{-i\vec p \cdot
         \vec r}}{(2\pi)^{3/2}} \,\,,
\end{equation}
where $\mathcal{A}$ is an antisymmetrization operator 
acting on the constituents
of both fragments, $\psi_1$ and $\psi_2$ are the internal wave functions of 
the two fragments, while $\vec p$ is the relative momentum and $\vec r$ the 
relative coordinate of the two fragments. Note that internal coordinates and
quantum numbers of $\psi_1$ and $\psi_2$ are suppressed in (\ref{plane}).

Substituting (\ref{lippmann}) into (\ref{ti1}) one finds that the transition 
matrix element is described by the sum of two pieces, a Born term
\begin{equation}\label{born}
   T_{\alpha k m}^{Born}(E_f)=\langle \phi_{\alpha}^{PW}|{\hat O}_{k}
                              |\psi_{0,m}\rangle 
\end{equation}
and a final state interaction (FSI) term
\begin{equation} \label{fsi}
  T_{\alpha k m}^{FSI}(E_f)=\langle \phi_{\alpha}^{PW}|\hat{V}_{12}\frac{1}
  {E_f+i\epsilon - \hat{H}} \, {\hat O}_{k}|\psi_{0,m}\rangle \mbox{.}
\end{equation}
While the Born term can be calculated without major problems the FSI term is 
more difficult. Using the completeness of the eigenstates 
$|\psi_p(E_p)\rangle$ of $\hat H$ one gets for the FSI term 
\begin{eqnarray}
   T_{\alpha k m}^{FSI}(E_f) &=& 
   \int df'\langle \phi_{\alpha}^{PW}|\hat{V}_{12}|\psi_{f'}(E_{f'})\rangle 
   \langle  \psi_{f'}(E_{f'})|\frac{1}{E_f+i\epsilon -\hat{H}}
               \,{\hat O}_{k}|\psi_{0,m}\rangle\\
  &=& \int df' \langle \phi_{\alpha}^{PW}|
              \hat{V}_{12}|\psi_{f'}(E_{f'})\rangle \langle 
  \psi_{f'}(E_{f'})|{\hat O}{_{k}}|\psi_{0,m}\rangle \frac{1}{E_f+i\epsilon -
  E_{f'}}\\
  &=& -i\pi F_{\alpha k m}(E_f)+{\mathcal{P}}\int_{E_0-\delta}^{\infty} dE_{f'}
\frac{F_{\alpha k m}(E_{f'})}{E_f-E_{f'}}
\end{eqnarray}
with 
\begin{equation}
\label{ffi}
F_{\alpha k m}(E_{f'})= \sum_{f'} \langle \phi_{\alpha}^{PW}|\hat{V}_{12}|
\psi_{f'}(E_{f'})
\rangle \langle \psi_{f'}(E_{f'})|{\hat O}_{k}|\psi_{0,m}\rangle \mbox{,}
\end{equation}
where the sum over $f'$ stands for a possible degeneracy of the states 
$|\psi_{f'}\rangle$ with energy $E_{f'}$.

Equation (\ref{ffi}) is only a formal solution for $F_{\alpha k m}(E)$, 
since one should know all the eigenstates of $\hat{H}$. A direct calculation
of $F_{\alpha k m}(E)$ is even more difficult than the original problem of 
Eq.~(\ref{ti1}), where one needs to know just the eigenstate of energy $E_f$. 
However, one obtains $F_{\alpha k m}$ indirectly applying the LIT approach:

\begin{eqnarray}
\label{loree}
L_{\alpha k m}(\sigma) &=&\int_{E_0-\delta}^{\infty}dE \frac{F_{\alpha k m}
(E)}{(E-\sigma)(E-\sigma^*)}\\
      &=&\int_{E_0-\delta}^{\infty} df' \langle \phi_{\alpha}^{PW}|
\hat{V}_{12}\frac{1}{\hat{H}-\sigma}|\psi_{f'}\rangle \langle \psi_{f'}|\frac{1}
{\hat{H}-\sigma^*}{\hat O}_{k}|\psi_{0,m}\rangle \\
      &=&  \langle \phi_{\alpha}^{PW}|\hat{V}_{12}\frac{1}{\hat{H}-\sigma}
\frac{1}{\hat{H}-\sigma^*}{\hat O}_{k}|\psi_{0,m}\rangle \\
      &=& \langle \widetilde{\psi}_{\alpha}(\sigma)|\widetilde{\psi}_{km}
(\sigma)\rangle \,.
\end{eqnarray}
The Lorentz state $\widetilde{\psi}_{km}(\sigma)$ is obtained solving 
essentially the same differential equation as for the inclusive
case discussed in the previous chapter:
\begin{equation} \label{psi11}
   (\hat{H}-\sigma_R+i\sigma_I )|\widetilde{\psi}_{km}\rangle ={\hat O}_{k}|
   \psi_{0,m}\rangle 
   \mbox{,}
\end{equation}
whereas $\tilde{\psi}_{\alpha}(\sigma)$ is the solution of new differential 
equation
\begin{equation} \label{psi22}
  (\hat{H}-\sigma_R+i\sigma_I )|\widetilde{\psi}_{\alpha}\rangle =\hat{V}_{12}|
  \phi_{\alpha}^{PW}\rangle 
  \mbox{.}
\end{equation}
If $\hat{V}_{12}$ represents a finite range potential, the term on the 
right hand
side of (\ref{psi22}) vanishes for large distances, and one has also for 
$\widetilde{\psi}_{\alpha}$ an asymptotic bound state boundary condition 
similar to a bound state. If $ \hat V_{12}$ is not of finite range one can 
reformulate the LIT method accordingly \cite{Efros99,Lapiana}.

Inverting the various $L_{\alpha k m}$ one gets the functions 
$F_{\alpha k m}$ and thus the transition matrix elements 
$T_{\alpha k m}^{FSI}$. An additional calculation of the Born terms
$T_{\alpha k m}^{Born}$ leads to the searched structure functions.   
Also for exclusive cross sections one obtains reliable results with the LIT
method as shown in a test calculation for the $d(e,e'N)$ reaction 
\cite{Lapiana}.

Now we turn to the application of the Lanczos method to the LIT of the 
exclusive reaction. To this end we first rewrite the LIT in the following form
\begin{equation} \label{loreaga}
  L_{\alpha k m}(\sigma)
      =\frac{1}{2i\sigma_I} \langle \phi_{\alpha}^{PW} | 
       \hat{V}_{12} 
       (\frac{1}{\hat{H}-\sigma_R-i\sigma_I}
       -\frac{1}{\hat{H}-\sigma_R+i\sigma_I}) 
       \hat{O}_k|\psi_{0,m} \rangle  \mbox{.}
\end{equation}
Using the completeness relation over the Lanczos 
vectors $\{ |\phi_i\rangle \mbox{,}i=0,\ldots,n\} $, where the starting vector 
$|\phi_0\rangle$ is the same as in (\ref{start}), and setting $z=\sigma_R+
i\sigma_I$ one finds
\begin{eqnarray}
L_{\alpha k m}(\sigma) &=& 
          \frac{i}{2\sigma_I} \{ \sum_{i=0}^{n} \langle 
          \phi_{\alpha}^{PW}| \hat{V}_{12}|\phi_i \rangle 
          [ \langle \phi_i|\frac{1}{z-\hat{H}}\hat{O}_k|\psi_{0,m}
            \rangle 
          - \langle \phi_i|\frac{1}{z^*-\hat{H}}\hat{O}_k|\psi_{0,m}
            \rangle ]\} \cr
         &=& \frac{i \sqrt{\langle \psi_{0,m} |\hat{O}_k^{\dagger}\hat{O}_k|
                      \psi_{0,m} \rangle}}{2\sigma_I} 
          \{ \sum_{i=0}^{n} \langle \phi_{\alpha}^{PW}| 
             \hat{V}_{12}|\phi_i \rangle 
          [ \langle \phi_i|\frac{1}{z-\hat{H}}|\phi_0\rangle - 
            \langle \phi_i|\frac{1}{z^*-\hat{H}}|\phi_0\rangle ]\}  \cr
         &=&  \frac{i\sqrt{\langle \psi_{0,m} |\hat{O}_k^{\dagger}\hat{O}_k|
              \psi_{0,m}\rangle}}{2\sigma_I} 
           \{ \sum_{i=0}^{n} \langle \phi_{\alpha}^{PW}| 
              \hat{V}_{12}|\phi_i \rangle [ x_{i0} - \widetilde{x}_{i0}]\} \,,
\end{eqnarray}
with the matrix elements
\begin{eqnarray}
   x_{0i} &=& \langle \phi_0|\frac{1}{z-\hat{H}}|\phi_i\rangle \mbox{,}\\
   \widetilde{x}_{0i} &=& 
   \langle \phi_0|\frac{1}{z^{*}-\hat{H}}|\phi_i\rangle \,,
\end{eqnarray}
which can be written as continued fractions of the Lanczos coefficients,
like in (\ref{conti}).

In the following we describe a simple algorithm for a recursive calculation
of all the $x_{0i}$ starting from $x_{00}$. Defining
\begin{equation}
   g(\nu)=-\frac{b_{\nu}^2}{z-a_{\nu}
          -\frac{b_{\nu+1}^2}{z-a_{\nu+1}
          -\frac{b_{\nu+2}^2}{\cdots}}}
\end{equation}
one has
\begin{equation} \label{xoo}
  x_{00}=\frac{1}{z-a_0+g(1)} \,.
\end{equation}
The next matrix element $x_{01}$ can be written as
\begin{equation} \label{xo1}
  x_{01}=\frac{1}{(z-a_1)\lambda_0-b_1+\lambda_0 g(2)} \,,
\end{equation}
where 
\begin{equation} \label{lambdao}
  \lambda_{0}=\frac{z-a_0}{b_1}
\end{equation}
is obtained from $x_{00}$.
Now a new parameter 
\begin{equation}
\label{lambda1}
\lambda_{1}=\frac{(z-a_1)\lambda_0-b_1}{b_2}  
\end{equation}
can be obtained from $x_{01}$ leading to the evaluation of
\begin{equation}
\label{xo2}
x_{02}=\frac{1}{(z-a_2)\lambda_1-b_2\lambda_0+\lambda_1 g(3)} \,.
\end{equation}
Proceeding in this way it is easy to show that the $n$-th matrix element 
$x_{0n}$ can be written as
\begin{equation}
\label{xon}
x_{0n}=\frac{1}{(z-a_n)\lambda_{n-1}-b_n\lambda_{n-2}+\lambda_{n-1}g(n+1)} 
\end{equation}
with
\begin{equation}
\label{lambdan}
\lambda_{n}=\frac{(z-a_n)\lambda_{n-1}-b_n\lambda_{n-2}}{b_{n+1}} \,.
\end{equation}

\section{Results}

In order to check the accuracy and to quantify the advantages of the Lanczos
method for the LIT approach we apply it to the calculation of total 
photoabsorption cross sections of $^{3}$H and $^{4}$He and compare 
to results which we obtain solving Eq.~(\ref{psi1}) via an inversion of the 
hamiltonian matrix. 

For the total photoabsorption cross section we use the dipole approximation,
where one has a single transition operator, namely 
\begin{equation}
{\hat O}_k = {\hat O}_{k'}=
\sum_i^A z_i \frac{1+\tau_z(i)}{2} \,;
\end{equation}
here $z_i$ and $\tau_z(i)$ are
the third components of the spatial coordinate and of the isospin of particle 
$i$, respectively. We expand ground and Lorentz states 
on the hyperspherical harmonics basis making use of the recently developed 
EIHH approach \cite{EIHH}. As NN interaction model we employ the semirealistic
MTI-III potential \cite{MT}. We note in passing that calculations of the total 
photoabsorption cross section of $^3$H and $^4$He via the LIT already exist 
\cite{ELO97b,ELO97c,ELOT00}. 

In Fig.~1 we show the LIT for the triton photodisintegration choosing 
the isospin channel $T=1/2$ of the disintegrated three-body system. One sees
that there is an excellent agreement between both calculations. The
relative difference of both results is far below 0.1 $\%$. For the 
second isospin channel ($T=3/2$), not shown here, the comparison is of the same 
excellent quality. In Fig.~2 we show the LIT for the case of the 
$^{4}$He photodisintegration, where one has only a single isospin channel 
of the disintegrated four-body system ($T=1$) for the considered dipole 
transition. Again one sees a very good 
agreement comparing the results with and without application of the Lanczos 
method. One finds only tiny differences between both results. It is evident 
from Figs.~1 and 2 that the Lanczos method leads to highly reliable results.

The main advantage of the Lanczos technique lies in the enormous reduction of 
the cpu time. In particular one should note that it is not necessary to perform
all the Lanczos steps. In fact, checking the 
convergence of $L(\sigma)$ with respect to the number of Lanczos coefficients 
used in the continued fraction one finds the following: for
$\sigma_R \rightarrow \infty$ only the first Lanczos step is necessary, 
because the entire continued fraction is dominated just by $\sigma_R$. 
Also in the low $\sigma_R$ region one can expect that the continued 
fraction converges rapidly, since for $\sigma_R \rightarrow 0$  the 
dominating contribution to the term $1/(z-\hat{H})$ is given by the lowest
eigenvalue of $\hat H$, which is well approximated using only the first few 
Lanczos coefficients. Thus one may hope that in the intermediate $\sigma_R$ 
region a rather small number of Lanczos coefficients will be sufficient. 
In Fig.~3  we illustrate the convergence behaviour with respect to the 
number of Lanczos steps for the LIT of the $^{4}$He photodisintegration 
for a test case with 1372 basis states. 
As one can see the transforms are somewhat different from the converged one 
only if one takes less than $100$ Lanczos coefficients. The relative error  
is already smaller than $0.5\%$ for 150 coefficients and becomes rapidly
smaller with a further increasing number of Lanczos steps. One can conclude 
that $200 \div 300$ steps are completely sufficient in the studied 
case. Also due to the rapid convergence we obtain a reduction in cpu time of 
two orders of magnitude comparing with the conventional LIT calculation.

Applying the Lanczos method one has to pay attention to the loss of 
orthogonality of the Lanczos vectors with an increasing number of steps, since 
it could reduce the precision of the Lanczos coefficients.  We check the 
orthogonality calculating the scalar products of the $0$-th and the 
$i$-th Lanczos vectors. The results for the LIT case, illustrated in 
Fig.~3, are shown in Table \ref{dotp}. It is evident that after 300 
Lanczos steps, where one has a sufficiently converged LIT, one has still a 
rather good orthogonality. In fact the overlap between the 0-th
and the 300-th Lanczos vector is quite small (about $10^{-4}$).
However, in other cases one could encounter problems with the orthogonality
and thus one would need to reorthogonalize the Lanczos vectors.

In the following we summarize our results briefly. The LIT approach for the 
calculation of both inclusive and exclusive cross sections including 
complete final state interactions is reformulated in order to use the Lanczos 
algorithm. The Lanczos method allows one to 
calculate the LIT without complete inversion of the hamiltonian matrix. 
It leads to a huge simplification of the 
calculation reducing the needed cpu time enormously (about two orders of 
magnitude) and, even more important, paving the way for calculations with
large number of basis functions. We have tested the reliability of the method 
with great success for the LIT of the total photoabsorption 
cross sections of $^3$H and $^4$He. The high precision of the method 
together with the strong reduction of the cpu time opens the way to 
precise microscopic calculations of reaction cross sections also for
more complex few-body systems.
Indeed, very recently the new method has been applied to the first 
microscopic calculation of the total photoabsorption cross section of the A=6 
nuclei with complete final state interactions of the six-nucleon system 
\cite{BMBLO}.

\begin{table}
\begin{center}
\begin{tabular}{ll}
$i$  &  $\langle \phi_0| \phi_i\rangle$ \\
\hline
       1          &   1.3 10$^{-16}$  \\
      150         &   -2.4 10$^{-6}$  \\
      200         &   -1.2 10$^{-4}$  \\
      250         &   -9.5 10$^{-5}$  \\
      300         &   1.1 10$^{-4}$  \\
\end{tabular}
\end{center}
\caption{Scalar product between the 0-th and $i$-th lanczos vector.}

\label{dotp}
\end{table}

\begin{figure}
\caption{LIT of the $^{3}$H photoabsorption cross section for $T=1/2$ with 
conventional and Lanczos methods: absolute (a) and relative values (b)
setting the conventional LIT result equal to 1 ($\sigma_I=20$ MeV).}
\label{trizio1}
\end{figure}

\begin{figure}
\caption{LIT of the $^{4}$He photoabsorption cross section with 
conventional and Lanczos methods: absolute (a) and relative values (b)
setting the conventional LIT result equal to 1 ($\sigma_I=20$ MeV).}
\label{trizio2}
\end{figure}

\begin{figure}
\caption{Convergence of the LIT of Fig.2 with respect to the number $n$
of Lanczos coefficients: absolute (a) and relative values (b)
setting the LIT result with 1372 Lanczos coefficients equal to 1.
}
\label{elio1}
\end{figure}


\begin{thebibliography}{00}

\bibitem{ELO94} V. D. Efros, W. Leidemann, and G. Orlandini, Phys. Lett. B
{\bf  338} (1994) 130.
\bibitem{Sara} S. Martinelli et al., Phys. Rev. C {\bf 52} (1995) 1778.
\bibitem{ELO97a} V.D. Efros, W. Leidemann, and G. Orlandini, Phys. Rev. Lett.
{\bf 78} (1997) 432.
\bibitem{ELO97b} V.D. Efros, W. Leidemann, and G. Orlandini, Phys. Rev. Lett.
{\bf 78} (1997) 4015; \\
N. Barnea, V. D. Efros, W. Leidemann, and G. Orlandini, 
Phys. Rev. C {\bf  63} (2001) 057002.
\bibitem{ELO97c} V. D. Efros, W. Leidemann, and G. Orlandini, Phys.  Lett. B
{\bf  408} (1997) 1.
\bibitem{ELO99} V. D. Efros, W. Leidemann, and G. Orlandini,  
Few-Body Syst. {\bf  26} (1999) 251.
\bibitem{ELOT00} V. D. Efros, W. Leidemann, and G. Orlandini, E. L. Tomusiak, 
Phys.  Lett. B {\bf  484}, 223 (2000); Nucl. Phys. {\bf A689} (2001) 421c.
\bibitem{dagotto}
E. Dagotto, Rev. Mod. Phys. {\bf 66} (3), 763 (1994).
\bibitem{hallberg}
K.A. Hallberg, Phys. Rev. B {\bf 52} 9827 (1995).
\bibitem{kuehner}
T.D. K\"uhner and S.R. White, e-print cond-mat/9812372. 
\bibitem{Efros85} V.D. Efros, Sov. J. Nucl. Phys. {\bf 41} (1985) 949.
\bibitem{lanczos} C. Lanczos, J. Res. Nat. Bur. Stand. {\bf 45}, 255 (1950). 
\bibitem{fulde}
P. Fulde, {\em Electron Correlations in Molecules and Solids}, Springer Series 
in Solid State Physics {\bf 100} (Springer-Verlag, Berlin/Heidelberg/New York)
\bibitem{Efros99} V.D. Efros, Phys. Atom. Nucl. {\bf 62} (1999) 1833.
\bibitem{Lapiana} A. La Piana and W. Leidemann, Nucl. Phys. {\bf A677} (2000)
 424.
\bibitem{EIHH} N. Barnea, W. Leidemann, and G. Orlandini, Phys. Rev. C 
{\bf  61} (2000) 054001; Nucl. Phys. A {\bf 693} (2001) 565.
\bibitem{MT} R. A. Malfliet and J. A. Tjon, Nucl. Phys. {\bf A127} (1969) 161. 
\bibitem{BMBLO} S. Bacca, M.A. Marchisio, N. Barnea, W. Leidemann, and G.
Orlandini, e-print nucl-th/0112067

\end{thebibliography}
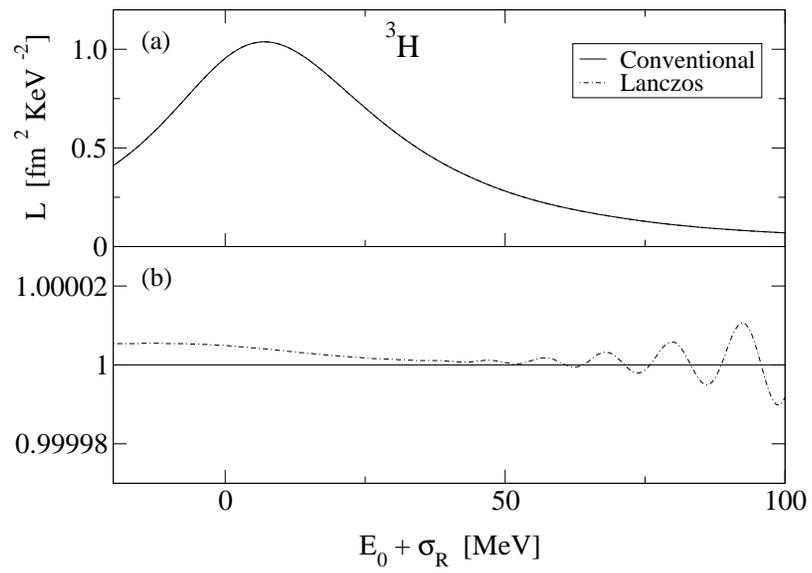
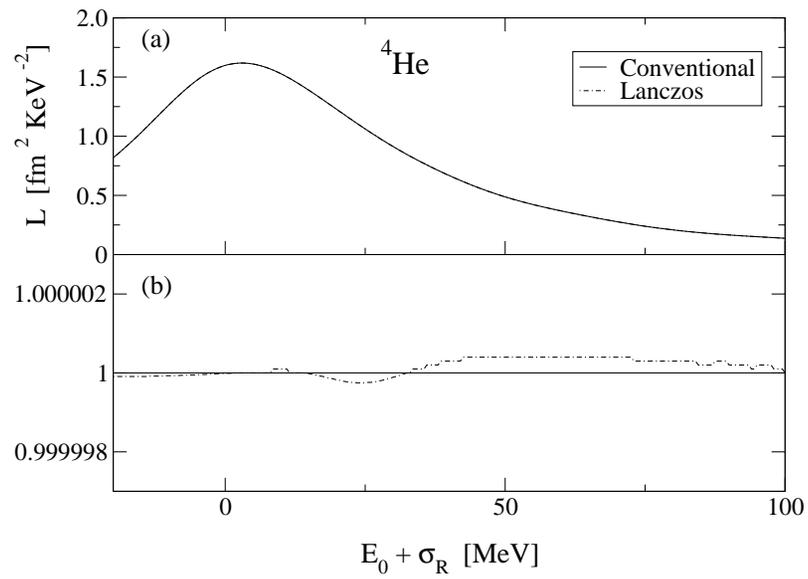
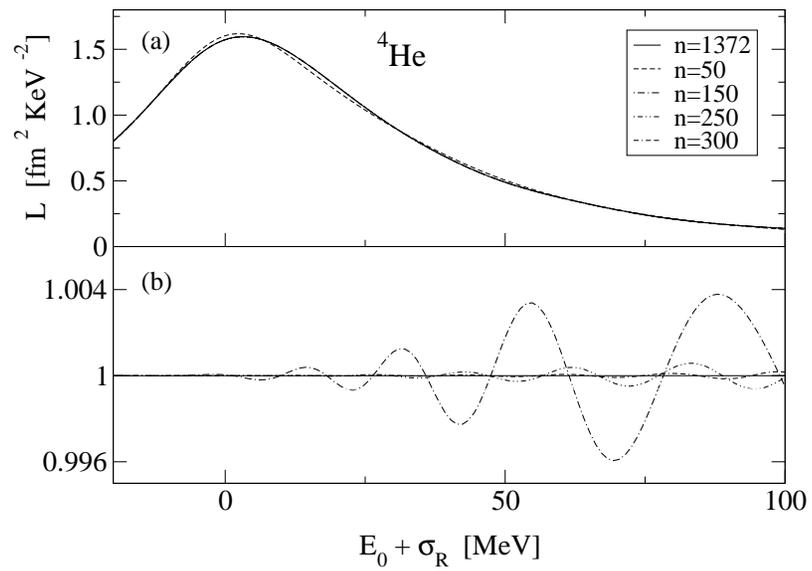
\end{document}